\documentclass[final]{svjour3}
\usepackage{graphicx}
\usepackage{rotating}
\usepackage{amssymb}
\usepackage{mathptmx}
\usepackage[numbers,sort&compress]{natbib}
\usepackage{lipsum}
\usepackage{subfig}
\usepackage{sidecap}
\usepackage{wasysym}
\usepackage[font={small}]{caption}
\usepackage{bm}
\usepackage{tabularx,ragged2e,booktabs}
\usepackage[hang,flushmargin,bottom]{footmisc} 
\makeatletter

\usepackage{mwe}
\usepackage{float}
\usepackage{setspace}
\newcolumntype{C}[1]{>{\Centering}m{#1}}

\newcommand\subparagraph{%
	\@startsection{subparagraph}{5}
	{\parindent}
	{3.25ex \@plus 1ex \@minus .2ex}
	{-1em}
	{\normalfont\normalsize\bfseries}}
\makeatother
\makeatletter
\def\@maketitle{\newpage
	\normalfont
	\vbox to0pt{\if@twocolumn\vskip-39pt\else\vskip-49pt\fi
		\nointerlineskip
		\makeheadbox\vss}\nointerlineskip
	\vbox to 0pt{\offinterlineskip\rubricwidth=\columnwidth
		\vskip-12.5pt
		\if@twocolumn\else 
		\divide\rubricwidth by144\multiply\rubricwidth by89 
		\vskip-\topskip
		\fi
		\hrule\@height0.35mm\noindent
		\advance\fboxsep by.25mm
		\global\advance\rubricwidth by0pt
		\rubric
		\vss}\vskip-42pt 
	\if@twocolumn\else
	\gdef\footnoterule{%
		\kern-3\p@
		\hrule\@width\columnwidth  
		\kern2.6\p@}
	\fi
	\setbox\authrun=\vbox\bgroup
	\hrule\@height 9mm\@width0\p@
	\pretolerance=10000
	\rightskip=0pt plus 4cm
	\nothanksmarks
	{\authorfont
		\setbox0=\vbox{\setcounter{auth}{1}\def\and{\stepcounter{auth} }%
			\hfuzz=2\textwidth\def\thanks##1{}\@author}%
		\setcounter{footnote}{0}%
		\global\value{inst}=\value{auth}%
		\setcounter{auth}{1}%
		\if@twocolumn
		\rightskip43mm plus 4cm minus 3mm
		\else 
		\fi
		\def\and{\unskip\nobreak\enskip{\boldmath$\cdot$}\enskip\ignorespaces}%
		\noindent\ignorespaces\@author\vskip3.62pt} 
		
	{\LARGE\bfseries
		\noindent\ignorespaces
		\@title \par}\vskip 1.62pt\relax 
		
	\if!\@subtitle!\else
	{\large\bfseries
		\pretolerance=10000
		\rightskip=0pt plus 3cm
		\vskip-5pt
		\noindent\ignorespaces\@subtitle \par}\vskip 0.24pt
	\fi
	\small
	\if!\@dedic!\else
	\par
	\normalsize\it
	\addvspace\baselineskip
	\noindent\@dedic
	\fi
	\egroup 
	\@tempdima=\headerboxheight
	\advance\@tempdima by-\ht\authrun
	\unvbox\authrun
	\ifdim\@tempdima>0pt
	\vrule width0pt height\@tempdima\par
	\fi
	\noindent{\small\@date\vskip -2.24pt} 
	\global\@minipagetrue
	\global\everypar{\global\@minipagefalse\global\everypar{}}%
	\vskip22.47pt
}
\makeatother
\usepackage{titlesec}
\let\subparagraph\relax 
\usepackage{titlesec}
\makeatletter
\journalname{Journal of Low Temperature Physics}
\bibpunct{}{}{,}{s}{}{,}

\usepackage{xspace} 

\def\spider{{\sc Spider}\xspace}

\def\planck{{\it Planck}\xspace}
\def\bicep{{\sc Bicep}\xspace}
\def\biceptwo{{\sc Bicep}2\xspace}
\def\mukcmbrts{$\mu$K$_{cmb}$$\sqrt{s}$}

\begin{document}

\title{SPIDER: CMB polarimetry from the edge of space}

\author{
R. ~Gualtieri$^1$ \and J.P. ~Filippini$^{1,2}$ 
\and P.A.R. ~Ade$^3$ \and M. ~Amiri$^4$ \and S.J. ~Benton$^5$ \and A.S. ~Bergman$^5$ \and R.~Bihary$^6$ 
\and J.J. ~Bock$^{7,8}$ \and J.R. ~Bond$^9$ \and S.A. ~Bryan$^{10}$ \and H.C. ~Chiang$^{11,12}$ 
\and C.R. ~Contaldi$^{13}$ \and O. ~Dor\'e$^{7,8}$ \and A.J. ~Duivenvoorden$^{14}$ \and H.K. ~Eriksen$^{15}$ 
\and M. ~Farhang$^{9,16}$ \and L.M. ~Fissel$^{16,17}$ \and A.A. ~Fraisse$^5$ \and K. ~Freese$^{14,18}$ 
\and M. ~Galloway$^{19}$ \and A.E. ~Gambrel$^5$ \and N.N. ~Gandilo$^{20,21}$ \and K. ~Ganga$^{22}$ 
\and R.V. ~Gramillano$^1$ \and J.E. ~Gudmundsson$^{14}$ \and M. ~Halpern$^4$ \and J. ~Hartley$^{19}$ 
\and M. ~Hasselfield$^{23}$ \and G. ~Hilton$^{24}$ \and W. ~Holmes$^8$ \and V.V. ~Hristov$^7$ \and Z. ~Huang$^9$ 
\and K.D. ~Irwin$^{25,26}$ \and W.C. ~Jones$^5$ \and C.L. ~Kuo$^{25}$ \and Z.D. ~Kermish$^5$ \and S. ~Li$^{5,16,27}$ \and P.V. ~Mason$^7$ \and K. ~Megerian$^8$ \and L. ~Moncelsi$^7$ \and T.A. ~Morford$^7$ \and J.M. ~Nagy$^{28,6}$ 
\and C.B. ~Netterfield$^{16,19}$ \and M. ~Nolta$^9$ \and B. ~Osherson$^1$ \and I.L. ~Padilla$^{16,20}$ 
\and B. ~Racine$^{15,29}$ \and A.S. ~Rahlin$^{30,31}$ \and C. ~Reintsema$^{24}$ \and J.E. ~Ruhl$^6$ 
\and M.C. ~Runyan$^8$ \and ~T.M. Ruud$^{15}$ \and J.A. ~Shariff$^9$ \and J.D. ~Soler$^{32,33}$ \and X. ~Song$^5$ 
\and A. ~Trangsrud$^{7,8}$ \and C. ~Tucker$^3$ \and R.S. ~Tucker$^7$ \and A.D. ~Turner$^8$ 
\and J.F. van der ~List$^5$ \and A.C. ~Weber$^8$ \and I.K. ~Wehus$^{15}$ \and D.V. ~Wiebe$^4$ 
\and E.Y. ~Young$^5$}

\institute{$^1$Department of Physics, University of Illinois at Urbana-Champaign, Urbana, IL, USA \\
	$^2$Department of Astronomy, University of Illinois at Urbana-Champaign, Urbana, IL, USA \\
	$^3$School of Physics and Astronomy, Cardiff University, CF24 3AA, UK \\
	$^4$Department of Physics and Astronomy, University of British Columbia, Vancouver, BC, Canada \\
	$^5$Department of Physics, Princeton University, Princeton, NJ, USA 08544, USA \\
	$^6$Physics Department, Center for Education and Research in Cosmology and Astrophysics, CaseWestern Reserve University, Cleveland, OH, USA \\
	$^7$Division of Physics, Mathematics and Astronomy, California Institute of Technology, Pasadena, CA, USA \\
	$^8$Jet Propulsion Laboratory, Pasadena, CA, USA \\
	$^9$Canadian Institute for Theoretical Astrophysics, University of Toronto, Toronto, ON, Canada \\
	$^{10}$School of Earth and Space Exploration, Arizona State University, Tempe, AZ, USA \\
	$^{11}$School of Mathematics, Statistics and Computer Science, University of KwaZulu-Natal, Durban, South Africa \\
	$^{12}$National Institute for Theoretical Physics (NITheP), KwaZulu-Natal, South Africa \\
	$^{13}$Blackett Laboratory, Imperial College London, London, UK \\
	$^{14}$The Oskar Klein Centre for Cosmoparticle Physics, Department of Physics, Stockholm University, Stockholm, Sweden \\
	$^{15}$Institute of Theoretical Astrophysics, University of Oslo, Oslo, Norway \\
	$^{16}$Department of Astronomy and Astrophysics, University of Toronto, Toronto, ON, Canada \\
	$^{17}$National Radio Astronomy Observatory, Charlottesville, NC, USA \\
	$^{18}$Department of Physics, University of Michigan, Ann Arbor, MI, USA \\
	$^{19}$Department of Physics, University of Toronto, Toronto, ON, Canada \\
	$^{20}$Department of Physics and Astronomy, Johns Hopkins University, Baltimore, MD, USA \\
	$^{21}$NASA Goddard Space Flight Center, Greenbelt, MD, USA \\
	$^{22}$APC, Univ. Paris Diderot, CNRS/IN2P3, CEA/Irfu, Obs de Paris, Sorbonne Paris Cite, France \\
	$^{23}$Pennsylvania State University, University Park, PA, USA \\
	$^{24}$National Institute of Standards and Technology, Boulder, CO, USA \\
	$^{25}$Department of Physics, Stanford University, Stanford, CA, USA \\
	$^{26}$SLAC National Accelerator Laboratory, Menlo Park, CA, USA \\
	$^{27}$Department of Mechanical and Aerospace Engineering, Princeton University, Princeton, NJ, USA \\
	$^{28}$Dunlap Institute for Astronomy \& Astrophysics, University of Toronto, Toronto, ON, Canada \\
	$^{29}$Harvard-Smithsonian Center for Astrophysics, Cambridge, MA, USA \\
	$^{30}$Fermi National Accelerator Laboratory, Batavia, IL, USA \\
	$^{31}$Kavli Institute for Cosmological Physics, University of Chicago, Chicago, IL, USA \\
	$^{32}$Max-Planck-Institute for Astronomy, Heidelberg, Germany \\
	$^{33}$Laboratoire AIM, Paris-Saclay, CEA/IRFU/Sap-CNRS-Universit\'e Paris Diderot, Gif-sur-Yvette Cedex, France
	\email{rgualtie@illinois.edu}}

\maketitle

\begin{abstract}
\spider is a  balloon-borne instrument designed to map the polarization of 
the millimeter-wave sky at large angular scales. \spider targets the \textsl{B}-mode 
signature of primordial gravitational waves in the cosmic microwave background (CMB), with a focus on mapping a large sky area
with high fidelity at multiple frequencies. \spider's first long-duration
balloon (LDB) flight in January 2015 deployed a total of 2400 antenna-coupled 
Transition Edge Sensors (TESs) at 90~GHz and 150~GHz. 
In this work we review the design and in-flight performance of the 
\spider instrument, with a particular focus on the measured performance of 
the detectors and instrument in a space-like loading and radiation environment.  
\spider's second flight in December 2018 will incorporate payload upgrades 
and new receivers to map the sky at 285~GHz, providing valuable 
information for cleaning polarized dust emission from CMB maps.

\keywords{Cosmic microwave background, inflation, bolometers, transition edge sensors, polarimetry}

\end{abstract}

\section{Introduction}
Precision observations of the CMB have 
transformed our understanding of the universe's history and composition. 
The next frontier of this endeavor is the characterization of the exceedingly 
faint pattern of polarization in this radiation field, with an intensity 
more than an order of magnitude smaller than that of the CMB's well-known 
temperature anisotropies.  An accurate map of this polarization across 
the sky encodes information about the local conditions within the 
primordial plasma that is complementary to that available from the temperature. 
Most intriguingly, the presence of an odd-parity ``\textsl{B}-mode'' 
pattern of polarization at large angular scales would be an unambiguous 
signature that our universe began with an inflationary epoch of rapid expansion, 
thus providing information on fundamental physics at energy scales 
far beyond those achievable at accelerators~\cite{Dodelson_Astro2010:2009}.  
A believable detection (or exclusion) of this faint signal will require 
polarimeters of enormous sensitivity, exquisite control of polarized instrumental systematics, 
and clean separation of the CMB from galactic and atmospheric foregrounds. 
Numerous scientific teams are now developing and deploying terrestrial, 
balloon-borne, and satellite instruments to seek out this signature in the millimeter-wave sky.  
Current observational constraints limit the tensor-to-scalar ratio $r$ 
-- the power of primordial gravitational waves relative to that of 
the primordial density perturbations -- to be less than 0.07 at 
95\% confidence~\cite{BK2015,Planck2016}.

\begin{figure*}[t]
\begin{center}
\includegraphics[height=1.6in]{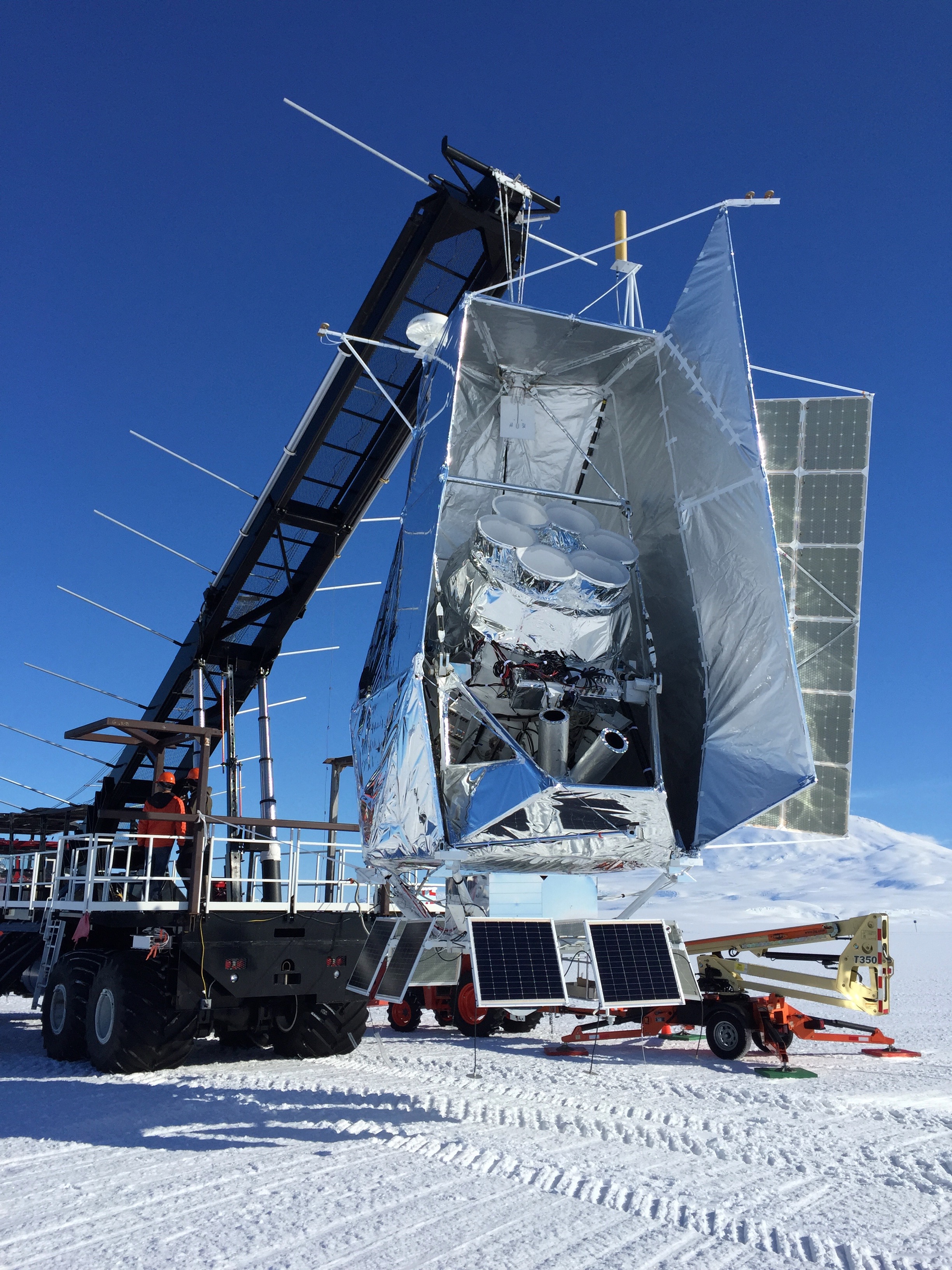}
\includegraphics[height=1.5in]{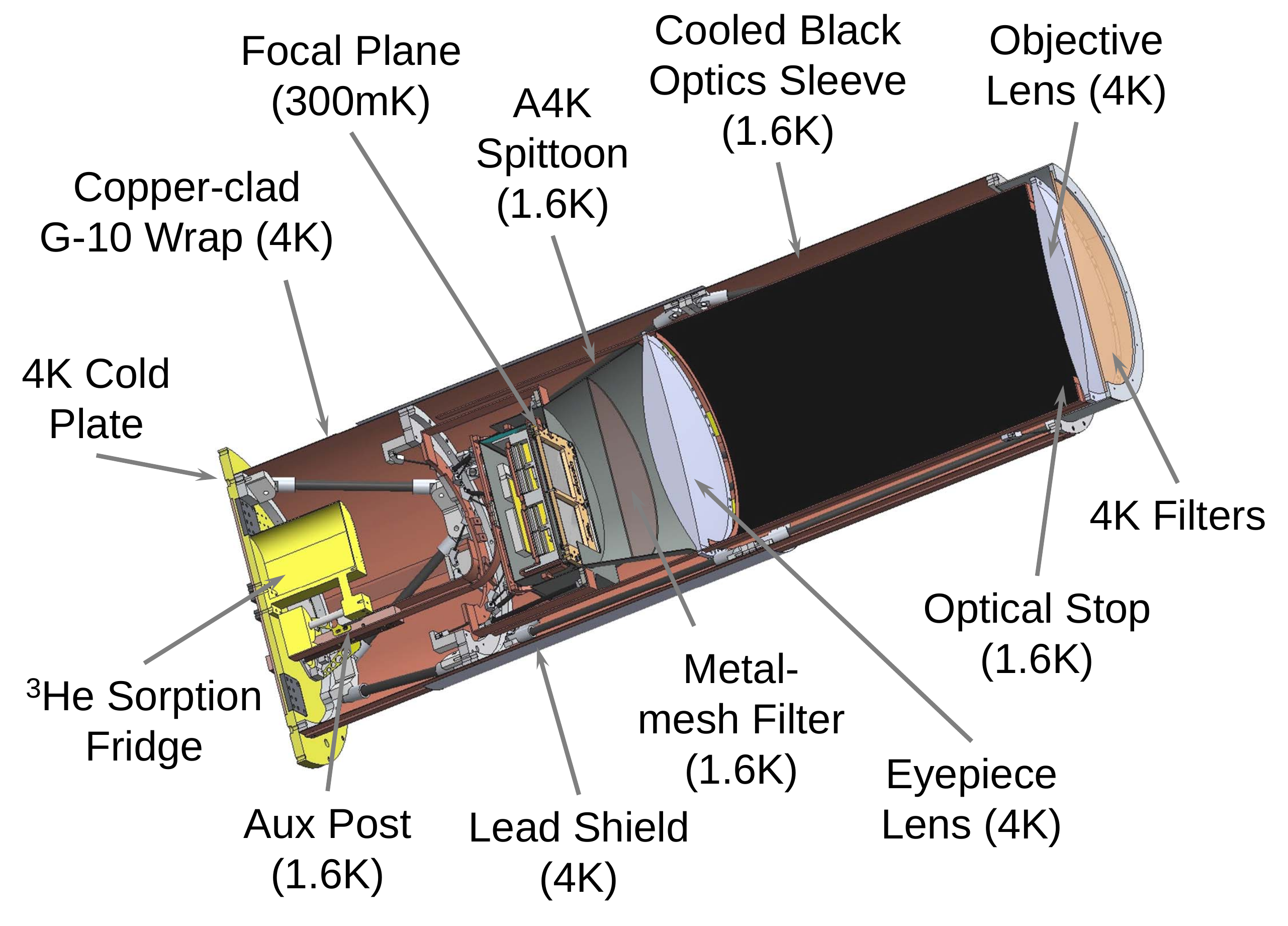}
\includegraphics[height=1.5in]{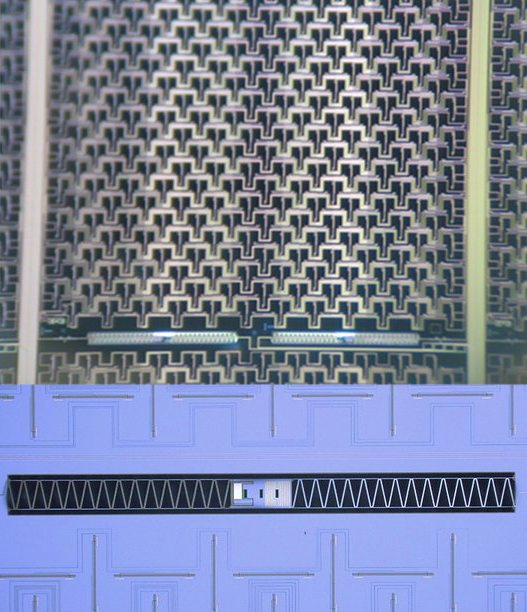}
\end{center}
\caption{{\it Left:} Photograph of the complete \spider payload hanging from the LDB launch vehicle. 
The six telescope aperture forebaffles are visible emerging from the central cryostat, which is supported by
a scanning carbon fiber gondola. A sun shield surrounds the assembly, and a solar panel ``wing'' is visible at right.
{\it Center:} Cross-section of the \spider telescope
  model with key components labeled. {\it Top right}: Antenna-coupled TES pixel as used in the first flight,
     showing phased array of slot antennas feeding two TESs at
     bottom. {\it Bottom right}: Close-up of a single TES assembly,
     showing the TES island and the SiN thermal isolation legs.  The
     meandered leg design allows for low thermal conductance in a
     narrow geometry.}
\label{fig:instrument}
\end{figure*}

\spider~\cite{Filippini2010,Fraisse2013,Rahlin2014} is a balloon-borne 
telescope array optimized to search for the signature of cosmic inflation\cite{Guth1981} 
in the polarization of the CMB, and to discriminate a primordial 
signal from galactic foreground emission.  \spider employs multiplexed 
arrays of transition-edge sensor (TES) bolometers to observe the millimeter-wave
sky from a NASA LDB. LDB flights from McMurdo Station, 
Antarctica, provide multi-week flights at high ($\sim$36~km) observing altitudes, 
near-space photon backgrounds, and sensitivity to frequencies and 
angular scales that are contaminated by atmosphere at even the most 
favorable ground-based observing sites.  In addition to their observing advantages, 
LDB flights are also technological proving grounds for future space missions~\cite{bock_epic08}. 
In this paper we briefly review the design and performance of the \spider instrument, 
with a particular focus on \spider's performance in this environment 
during its first observing flight in January 2015.

\section{The \textbf{\textsc{Spider}} 2015 payload}

\subsection{Overview}
The \spider payload (Fig.~\ref{fig:instrument}, \textit{left}) consists 
of six monochromatic refracting telescopes housed within a single 
large liquid helium cryostat.  The cryostat and associated electronics 
are supported within a lightweight carbon fiber gondola, which hangs beneath the balloon~\cite{Soler2014}.  


A reaction wheel and linear drive scan the cryostat in azimuth and 
elevation throughout the flight working in concert with a motorized pivot, 
at speeds as high as $4 ^\circ/$s in azimuth~\cite{Shariff2014}.  
A suite of star cameras, GPS receivers, sun sensors, and gyroscopes 
enable accurate instantaneous ($<5'$~rms) and post-flight ($<5''$~rms) 
pointing reconstruction~\cite{Gandilo2014}, well in excess of requirements
 for the relatively broad beams ($30'$ FWHM at 150~GHz and $42'$ FWHM at 90~GHz)
  of the \spider telescopes.  A set of sun shields protects the instrument 
  and optics during the 24-hour Antarctic summer daylight.  
  A 2~kW solar panel system provides electric power throughout the flight.  
  An assortment of antenna arrays provide commanding, telemetry, and location information during flight.  

\spider's cryogenic system, the largest yet deployed on an LDB, 
consists of two liquid helium reservoirs: a 1284~L main tank and a 
20~L superfluid tank~\cite{Gudmundsson2015}.  The main tank is maintained 
near sea level atmospheric pressure during the flight, providing each telescope 
with cooling power at $\sim$4~K.  Vapor from the main tank 
cools a series of vapor-cooled shields to reduce radiative load on the cryogenic system.  
The superfluid tank fills continuously from the main tank through a capillary assembly, 
and is pumped out to about 5~torr by opening a valve to the low pressure atmosphere at float. 
The superfluid system provides cooling power at 1.6~K to each inserts' 
sub-Kelvin cooler and internal optical baffles.  
The focal planes themselves are cooled to $\sim$300~mK by a 
dedicated $^3$He adsorption refrigerator within each telescope.

\subsection{Receivers}
Each \spider receiver is a two-lens cold refracting telescopes, adapted from 
the basic plan of the \bicep1 instrument~\cite{Keating2003,Yoon2006}.
The \spider lens design is optimized for a planar-antenna focal plane, while 
the filter stack and telescope structure incorporate 
several adaptations for a balloon-borne environment.  
Each telescope consists of two anti-reflection coated polyethylene 
lenses cooled to 4~K, focusing light onto 
the focal plane (Fig.~\ref{fig:instrument}, \textit{center}).  
This compact, axisymmetric design limits polarized systematics 
within the instrument.  The relatively small (26~cm \diameter) 
stop aperture simplifies baffling and far-field beam characterization 
while retaining sufficient angular resolution to map the inflationary 
\textsl{B}-mode signal expected at degree angular scales.  
This modular design using monochromatic telescopes greatly simplifies 
anti-reflection coatings for the optical elements and allows for 
flexibility in the instrument's frequency coverage.  
Each receiver is supported by a rigid carbon fiber frame to reduce payload mass, 
a critical concern for balloon-borne operations.

A primary concern for the \spider optical system has been to reduce 
stray optical load on the detectors, which degrades mapping speed and 
would eventually saturate the TES bolometers.  This has forced several 
adaptations from the \biceptwo design~\cite{Bicep2instrument}.  
A blackened ``sleeve'' baffle cooled to $\lesssim$2 Kelvin by the 
superfluid tank has been added between the secondary lens and the 
Lyot stop to reduce loading from the telescope itself.  \spider's 
filter stack is predominantly reflective (metal-mesh filters~\cite{AdeMetalMesh}) 
to further reduce emission, with a single absorptive nylon filter at 4~K.  
The vacuum seal for each telescope is provided by a thin sheet of 
anti-reflection coated ultra-high molecular weight polyethylene, 
with low scattering and emissivity.  The warm baffling of each 
telescope is primarily reflective rather than absorptive, redirecting 
side lobe response to the cold sky rather than absorbing it at ambient temperature.

Polarization modulation for each telescope is provided by a sapphire 
half-wave plate~\cite{Bryan2010,Bryan2016}, just skyward of the 
4~Kelvin filter stack and also cooled to 4~K.  Each plate's 
thickness and coating are tuned for the wavelength of the associated 
receiver, yielding minimal impact on optical throughput, ghosting, and pointing.  
Each wave plate is operated in a stepped configuration, rotated 
every $\sim$12 hours during fridge cycling to provide polarization 
modulation over the course of the flight.

\subsection{Focal plane assemblies}
In \spider's 2015 flight, these optical systems focused light onto 
planar arrays of dual-polarization antenna-coupled TESs developed at Caltech/JPL~\cite{BKSdet2015}.  
Each receiver's focal plane houses four silicon tiles, each photolithographically 
patterned with an $8\times8$ square array ($6\times6$ for 90~GHz) 
of polarimeter pixels.  Each of these consists of two interpenetrating 
arrays of slot antennas sensitive to perpendicular polarization 
modes (Figure~\ref{fig:instrument}, \textit{top right}).  
The power received by each beam-forming antenna array is fed to a 
dedicated Ti TES ($T_c\sim500$ mK) through an inline band-defining filter.  
The arrays are read out using a time-division Superconducting 
QUantum Interference Device (SQUID) multiplexer system~\cite{dekorte_squid_mux,Battistelli2008mce}.  
A second Al TES ($T_c\sim1.3$ K) is wired in series with the 
primary Ti TES to allow receiver characterization under laboratory loading conditions.

\spider's detectors are optimized to take advantage of the extremely 
low optical loading from the sky at balloon altitudes, $<1/10$ that available 
at the South Pole.  \spider TESs use a meandered leg design 
(Fig.~\ref{fig:instrument}, \textit{bottom right}) that gives lower 
thermal conductance (and thus lower thermodynamic noise) in a 
similar footprint area.  Typical thermal conductances of the 
completed \spider arrays are of order $G\sim11$-$20$~pW/K at 450~mK. 
This solution is different in example from the straight-legged 
design used for the isolation membranes of the \biceptwo bolometers~\cite{Bicep2Antennas}.  
The electrical noise of the amplifier is kept subdominant to 
that from the detectors and photons by reducing the detector 
resistance and increasing the readout circuit inductance.

The focal plane housing the detectors and SQUIDs has been designed 
from the ground up to provide excellent shielding against time-varying 
magnetic fields, using several nested layers of superconducting 
and ferromagnetic materials.  Without such shielding the payload's 
motion in the Earth's magnetic field could induce spurious signals 
into the readout system.  This is a greater concern for \spider 
than for terrestrial instruments, since the gondola's motion makes 
it difficult to subtract such effects with a ground-fixed template. 
For further details of this design, see Runyan et al. 2010~\cite{Runyan2010}.

\section{In-flight performance}

\subsection{2015 flight}
\spider took to the air from the LDB base camp at McMurdo Station, 
Antarctica on January 1, 2015. All systems functioned well during 
the 16-day flight, with the exception of a redundant differential GPS package.  
The flight terminated successfully in West Antarctica, $2270$ km from the launch point. 
\spider's data disks were recovered in February 2015, and the 
full payload in November 2015, by the British Antarctic Survey.  
All hardware has been returned to North America, and is 
currently being refit for \spider's upcoming second flight (see Section 4).


Robust automation was a key consideration for \spider's flight operations. 
The limited satellite link bandwidth prohibited a live downlink of \spider's
full data set, necessitating on-board data quality checks and error recovery.
Payload telemetry was also limited, 
due both to bandwidth demands and to radio-frequency 
interference (RFI) induced in the detectors by the onboard higher-bandwidth antennas.  
Instead, the redundant flight computer systems monitored payload 
and detector status onboard, adjusting detector biases and 
other settings as needed. Highly compressed diagnostic 
packets were transmitted to the ground station in brief bursts 
at regular intervals, limiting the impact on data.  
The higher bandwidth satellite links were reserved for 
on-demand use to examine specific issues, generally during 
cryogenic servicing operations when science data were not being acquired.  

During its flight \spider mapped approximately 10\% of the sky 
at 90 and 150 GHz, using three receivers at each frequency.  
The resulting $\sim$1.6~TB of data are currently under analysis~\cite{Nagy2017}.

\subsection{Detector performance}
The 2015 flight incorporated a total of 2400 TES bolometers, 
of which 96 were intentionally not coupled to antennas.  
After accounting for hardware yield and relatively stringent data 
selection cuts, we currently retain data from 1863 optically-coupled 
bolometers (675/1188 at 90/150~GHz).  We further exclude periods of 
data on individual detectors for various causes, most prominently 
RFI from the communications arrays.  The estimated noise-equivalent 
temperature (NET) of the total instrument after these cuts is 
approximately 7.1~(5.3)~\mukcmbrts ~at 90~(150)~GHz.

Due to \spider's cold optical chain, radiation loading from the instrument 
itself was kept extremely low.  We estimate a total in-band 
loading during flight of $\lesssim~0.25$ ($\lesssim~0.35$)~pW at 90~(150)~GHz. 
This includes loading from the instrument, the atmosphere, and 
the CMB itself. Those numbers are comparable to, or lower 
than, those achieved by \planck-HFI at L2~\cite{PlanckHFIdesign}, a 
significant technical achievement and benefit of the LDB platform. 
If outfitted with truly photon-noise-limited detectors, 
a similar design might thus quadruple \spider's present mapping speed.

\subsection{Cosmic ray response}
One important challenge of a space or near-space environment is 
the greatly increased flux of particle radiation relative to that at sea level.  
The near-Earth environment is pelted by radiation of solar, galactic, and 
extra-galactic origins, predominantly a broad spectrum of protons 
peaking near 300~MeV and extending well beyond the TeV range.
Frequent glitches in the data stream from particle interactions 
have been a source of analysis difficulty for the \planck 
satellite~\cite{Planck2013_X_CosmicRay}, and will be a factor in 
the design of future space-based bolometric instruments.  
This particle flux is lower at balloon altitudes, but is still an important 
part of \spider's data analysis.  Below we give a first description 
of the general features of cosmic ray-induced glitches seen in the \spider data set.

Particle events deposit energy in the TES bolometer, producing brief 
glitches in the data that generally recover quickly according the bolometer's 
(optical) time constant. The typical energy deposition in a component is determined 
predominantly by its thickness, while the glitch rate is determined by its area. 
The incident particle rate is affected by solar activity, surrounding materials, 
and (for a balloon) the Earth's magnetic field. From a highly simplified model, 
we expect energy depositions of order $\sim$1~keV in each bolometer every 
few minutes, while the larger wafer experiences MeV-scale depositions at 
$\sim$100~Hz. 
The rate of bolometer depositions should thus be manageable in analysis, 
as long as the TES recovers quickly from each glitch. 
Of more concern are response to bolometer hits in the wafer and multiplexer 
crosstalk, both of which could greatly increase the data loss.


Cosmic ray-induced glitches (CRGs) are seen in \spider 2015 data 
every $\sim3$ minutes in each detector. Typical glitches are only a 
handful of samples in length at an acquisition rate of 126~Hz, 
with shape determined by the digital anti-aliasing filter applied 
by our readout electronics. These glitches are flagged and removed 
with minimal effect on our science analysis.  Laboratory 
characterization of an antenna-coupled TES array with a $^{32}$S 
source at Caltech revealed that the SQUID readout system could occasionally 
lose lock during a large energy deposition, causing the DC level 
of the data stream to be offset by one flux quantum. Such ``flux-slip'' 
discontinuities are seen regularly in \spider data, primarily due to 
RFI but occasionally due to energetic particle interactions.  
They are corrected during low-level data processing.

%

\begin{figure}[htpb]
	\centering
	\includegraphics[width=0.48\linewidth, keepaspectratio]{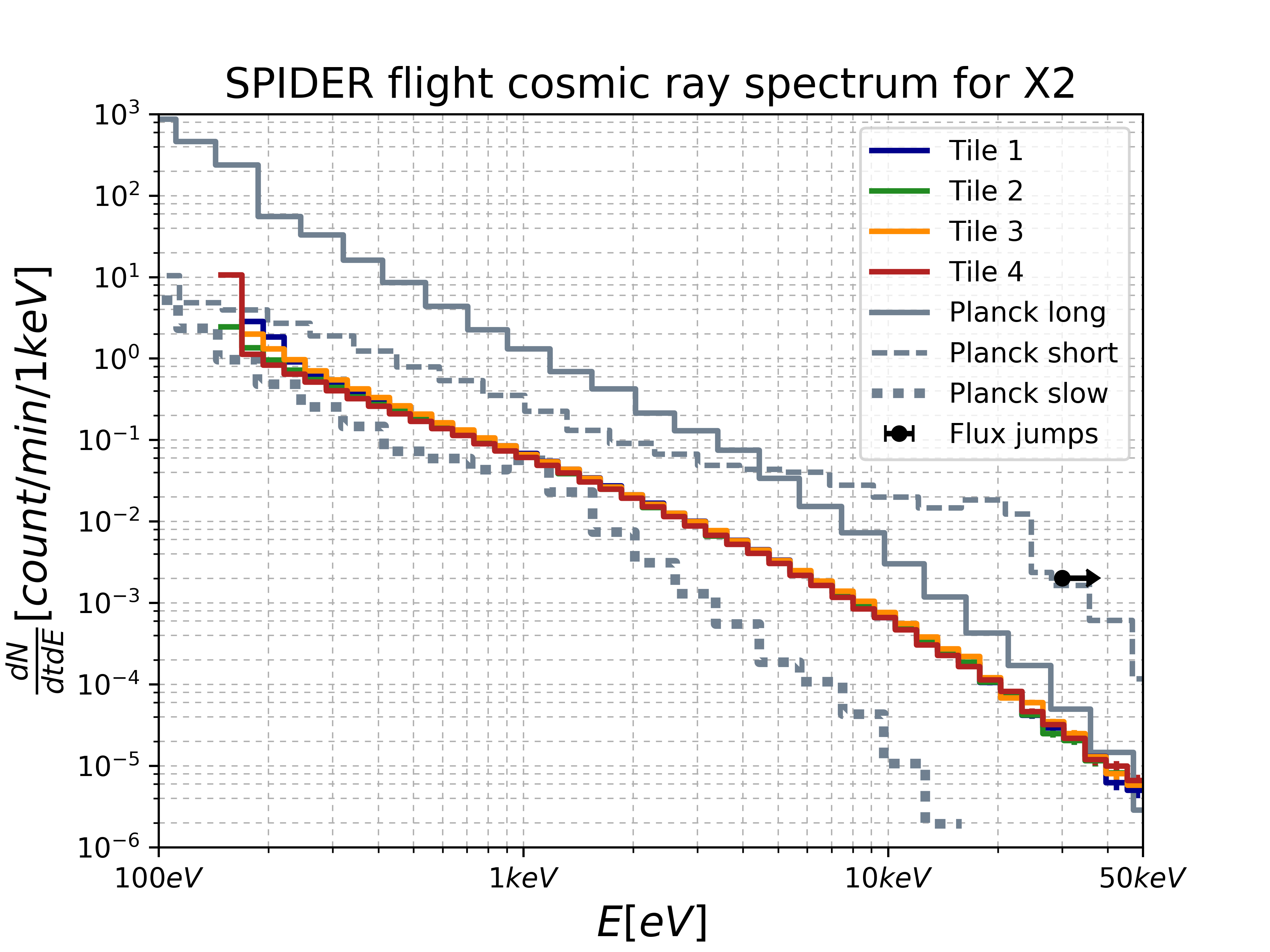}
	\includegraphics[width=0.48\linewidth, keepaspectratio]{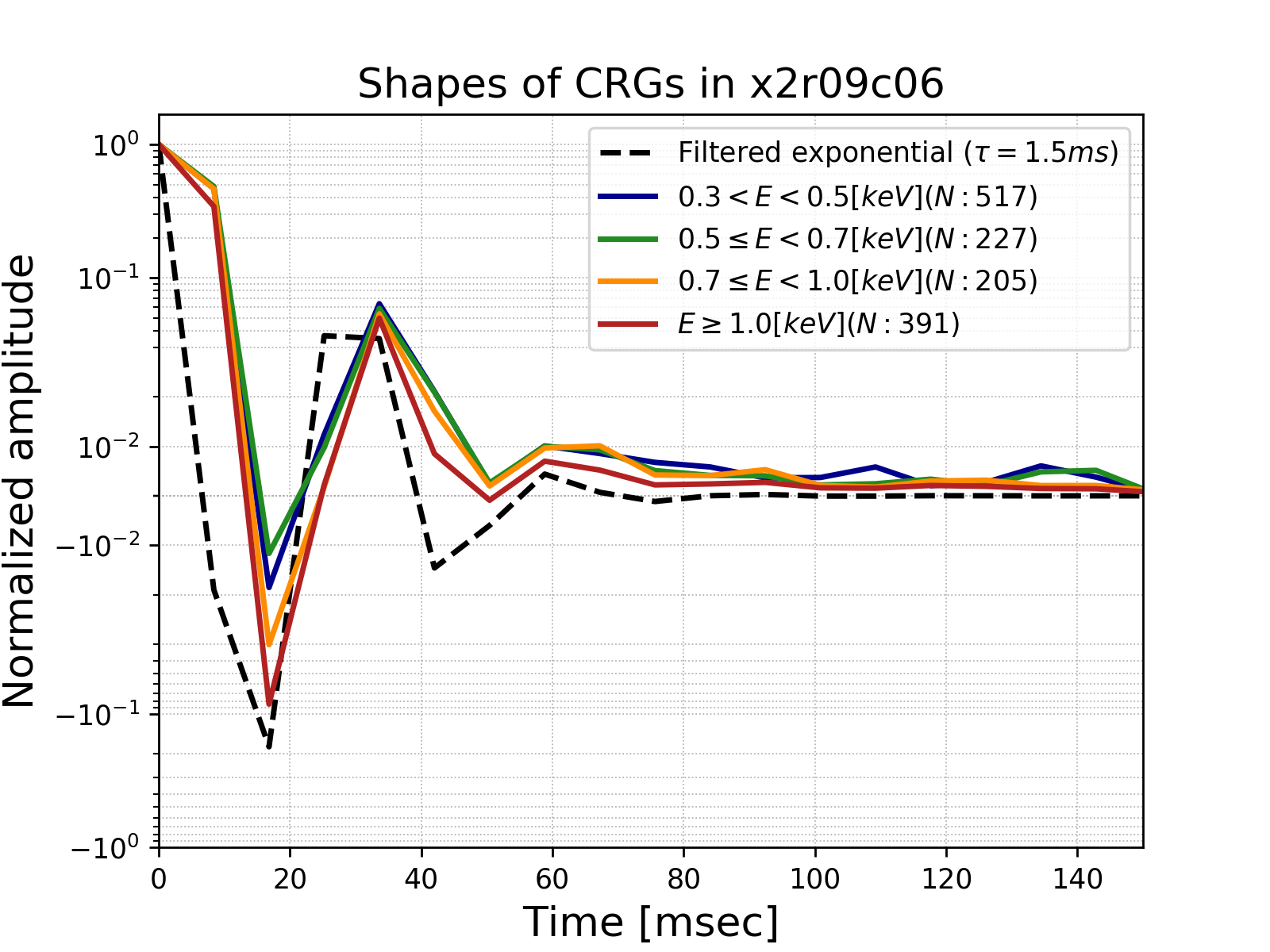}
	\caption{{\it Left}: Estimated spectrum of deposited energy for cosmic ray glitches in each of the four silicon tiles for a single \spider receiver. Approximate energies are estimated from the integrated reduction in Joule heating of the TES. Also plotted is the rate of flux slip discontinuities seen by this FPU (black marker near 30~keV) and spectra for the three types of CRG events seen in \planck \cite{Catalano2014}.
	{\it Right}: Averaged traces of CRGs seen in one representative detector.  The pulse shape is relatively independent of energy, and largely established by the digital anti-aliasing filter.}
	\label{fig:x2_spec_color}
\end{figure}


Fig.~\ref{fig:x2_spec_color} shows the estimated spectrum of deposited 
energies for cosmic ray glitches in a representative focal plane unit (FPU). 
Glitch rates are $10$-$100\times$ lower than in \planck HFI~\cite{Catalano2014}, 
qualitatively consistent with the somewhat more benign environment 
in the stratosphere. Coincidence rates are low ($\sim0.03\%$), 
suggesting relatively little long-distance energy propagation through the silicon tile. 
No long ``tails'' are apparent in the pulse decay times, but 
this is difficult to constrain given the digital filter.  
Laboratory tests with radioactive sources are in progress, 
and will give further information on this front.



\section{Status and prospects}
\begin{figure}[htpb]
	\centering
	\includegraphics[width=0.48\linewidth, keepaspectratio]{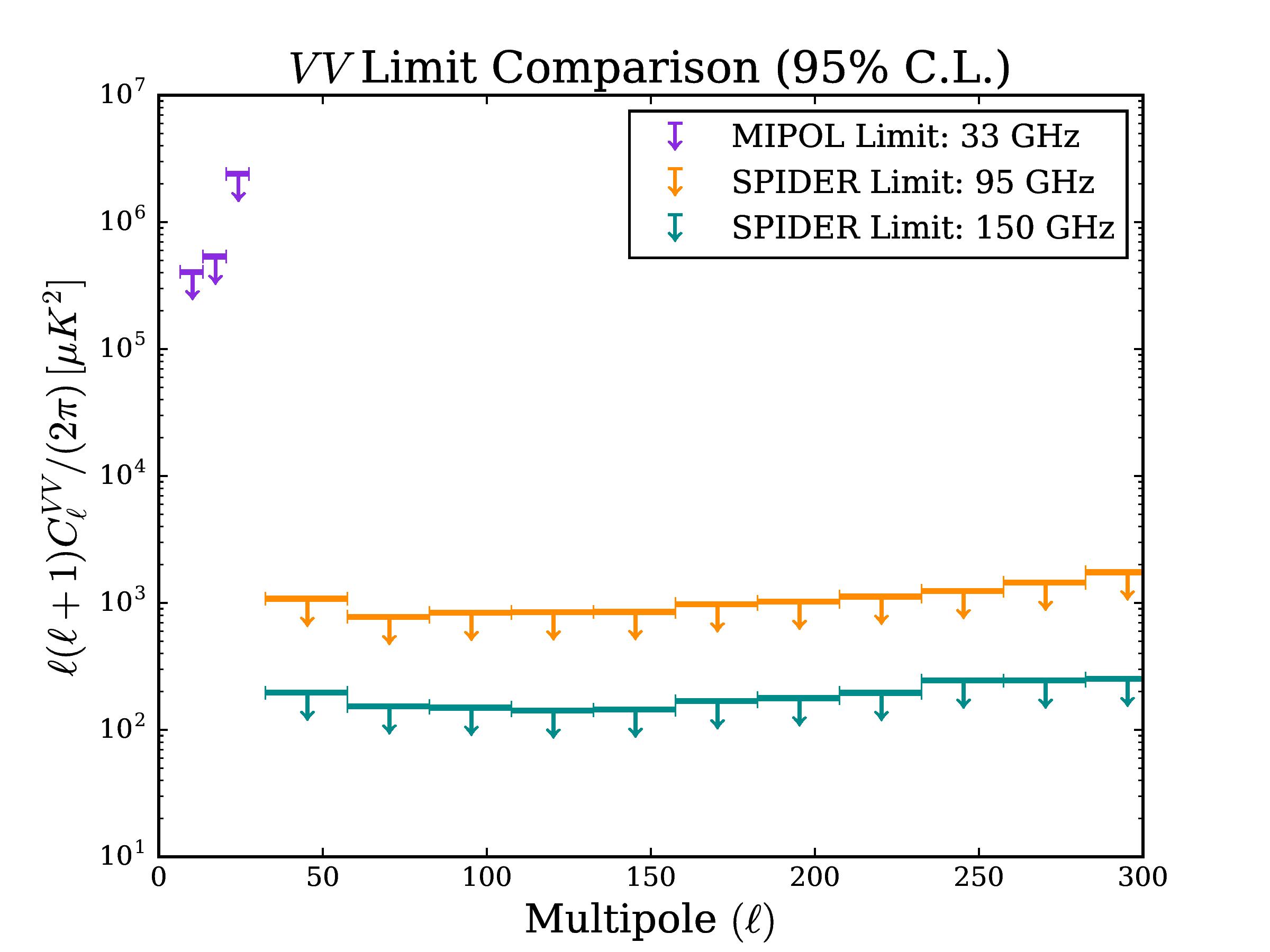}
	\includegraphics[width=0.51\linewidth, keepaspectratio]{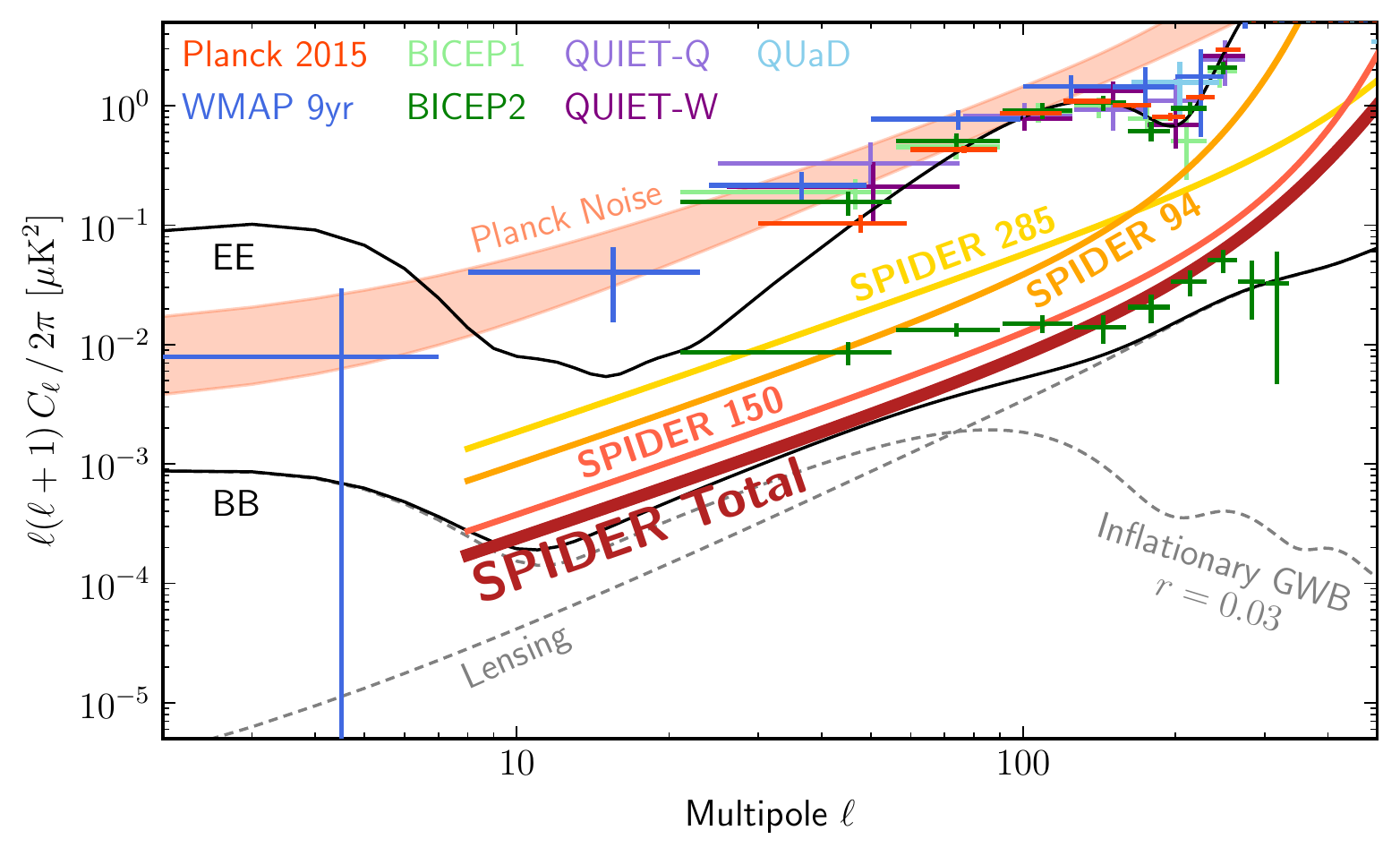}
	\caption{{\it Left}: Upper limits on circular polarization sky power at 90 and 150~GHz as a function of multipole, including comparison to previous results.  Figure from Nagy et al. (2017)~\cite{Nagy2017}.
	{\it Right}: Forecast BB sensitivity as a function of multipole for \spider after two flights, in comparison with some other recent experiments.}
	\label{fig:nell_apra2016} 
\end{figure}

\spider's first scientific result was published in 2017: 
a new constraint on the level of circular polarization in the 
CMB and millimeter-wave sky (Fig.~\ref{fig:nell_apra2016}, {\it Left}).  
This analysis takes advantage of non-idealities in \spider's half-wave 
plates that couple circular polarization into linear polarization, 
leading to constraints three orders of magnitude stronger than 
previous work~\cite{Nagy2017}.  \spider's primary B-mode polarization analysis is ongoing.

The \spider collaboration is preparing for a second observing flight 
in Dec. 2018, also on an Antarctic LDB.  This flight will incorporate 
three new receivers to map the same sky at 280~GHz, using feedhorn-coupled 
arrays of AlMn TES polarimeters developed at NIST~ \cite{Hubmayr2016,Bergman2017}.  
The resulting 280~GHz data will yield maps of galactic dust polarization 
over a large sky fraction at a frequency in between the existing \planck 
sky maps, which will have lasting value in the field.  \spider's second 
flight will also incorporate significant upgrades to the cryostat and gondola hardware.  
We expect \spider to be able to detect or constrain primordial 
gravitational waves at the $r>0.03$ at 3$\sigma$ confidence level (Fig.~\ref{fig:nell_apra2016}, {\it Right}). 
The results obtained in this work were possible thanks to the unique 
opportunity to study the CR environment from the stratosphere, 
for the first time with TESs, given by the data collected during 
the first successful \spider campaign.
The analysis presented here will help the design and deployment 
of this detector's technology on future satellite mission.

\begin{acknowledgements}
\spider is supported in the U.S. by the National Aeronautics and 
Space Administration under grants NNX07AL64G, NNX12AE95G, and NNX17AC55G 
issued through the Science Mission Directorate and by the 
National Science Foundation through PLR-1043515. Logistical support for the 
Antarctic deployment and operations was provided by the NSF through the U.S. Antarctic Program. 
Cosmic ray response studies are supported by NASA under grant 14-SAT14-0009. 
Support in Canada is provided by the National Sciences and 
Engineering Council and the Canadian Space Agency. Support in Norway 
is provided by the Research Council of Norway. Support in Sweden is 
provided by the Swedish Research Council through the Oskar Klein Centre 
(contract no. 638-2013-8993). KF acknowledges support from DoE grant 
DE-SC0007859 at the University of Michigan. The collaboration is 
grateful to the British Antarctic Survey, particularly Sam Burrell, 
for invaluable assistance with data and payload recovery after the 2015 flight. 
We also wish to acknowledge the generous support of the David and 
Lucile Packard Foundation, which has been crucial to the success of the project.
\end{acknowledgements}

\pagebreak

\end{document}